\documentstyle[12pt]{article}
\input amssym.def
\input amssym.tex

\newcommand{\wx}{\widetilde{x}}
\newcommand{\wip}{\widetilde{p}}
\newcommand{\wc}{\widetilde{c}}

\newcommand{\qed}{\rule{3mm}{3mm}}

\begin{document}

\def\m@th{\mathsurround=0pt}
\mathchardef\bracell="0365 
\def\upbrall{$\m@th\bracell$}
\def\undertilde#1{\mathop{\vtop{\ialign{##\crcr
    $\hfil\displaystyle{#1}\hfil$\crcr
     \noalign
     {\kern1.5pt\nointerlineskip}
     \upbrall\crcr\noalign{\kern1pt
   }}}}\limits}

\begin{titlepage}{\LARGE
\begin{center} New integrable systems \\
related to the relativistic Toda lattice
 \end{center}} \vspace{1.5cm}
\begin{flushleft}{\large Yuri B. SURIS}\end{flushleft} \vspace{1.0cm}
Centre for Complex Systems and Visualization, University of Bremen,\\
Kitzb\"uhler Str. 2, 28359 Bremen, Germany\\
e-mail: suris @ cevis.uni-bremen.de

\vspace{1.5cm}  
{\small {\bf Abstract.} New integrable lattice systems are introduced,
their different integrable discretization are obtained. B\"acklund 
transformations between these new systems and the relativistic Toda lattice 
(in the both continuous and discrete time formulations) are established.}
 
\end{titlepage}

\setcounter{equation}{0}
\section{Introduction}

This paper is devoted to integrable equations of the classical mechanics.
More precisely, we shall deal here with equations of motion in the Newtonian
form.

We introduce here two new integrable continuous time lattice systems, and 
present several novel integrable discrete time systems.

The first continuous time system is:
\begin{equation}\label{new1}
\ddot{x}_k=\dot{x}_{k+1}\exp(x_{k+1}-x_k)-\exp(2(x_{k+1}-x_k))-
\dot{x}_{k-1}\exp(x_k-x_{k-1})+\exp(2(x_k-x_{k-1})).
\end{equation}

The second one:
\begin{equation}\label{new2}
\ddot{x}_k=\dot{x}_k^2\Big(\dot{x}_{k+1}\exp(x_{k+1}-x_k)-
\dot{x}_{k-1}\exp(x_k-x_{k-1})
\Big).
\end{equation}

To the author's knowledge, these systems have not appeared in the literature,
despite their beauty and possible physical applications. However, they are
in a close relation to another well known integrable lattice -- 
a relativistic Toda lattice: 
\begin{equation}\label{old}
\ddot{x}_k=
\dot{x}_{k+1}\dot{x}_k\frac{g^2\exp(x_{k+1}-x_k)}{1+g^2\exp(x_{k+1}-x_k)}-
\dot{x}_k\dot{x}_{k-1}\frac{g^2\exp(x_k-x_{k-1})}{1+g^2\exp(x_k-x_{k-1})}.
\end{equation}
More precisely, we shall show that (\ref{new1}) and (\ref{old}) are connected
by a sort of B\"acklund transformation, and the same is true for (\ref{new2}) 
and (\ref{old}).

We now write down integrable discretizations we propose for the lattices 
(\ref{new1}), (\ref{new2}). In the difference equations below $x_k=x_k(t)$ are 
supposed to be functions of the discrete time $t\in h{\Bbb Z}$, and 
$\wx_k=x_k(t+h)$, $\undertilde{x_k}=x_k(t-h)$. Discretization of the first
lattice (\ref{new1}):
\vspace{0.4cm}

\noindent$\exp(\wx_k-x_k)-\exp(x_k-\undertilde{x_k})=$
\begin{equation}\label{dnew1}
-\frac{1}{1-h\exp(\undertilde{x_{k+1}}-x_k)}+h\exp(x_{k+1}-x_k)
+\frac{1}{1-h\exp(x_k-\wx_{k-1})}-h\exp(x_k-x_{k-1}).
\end{equation}

Discretization of the second lattice (\ref{new2})::
\vspace{0.4cm}

\noindent$\displaystyle\frac{h}{\exp(\wx_k-x_k)-1}-
\displaystyle\frac{h}{\exp(x_k-\undertilde{x_k})-1}=$
\begin{equation}\label{dnew2}
\exp(x_{k+1}-x_k)-\exp(\undertilde{x_{k+1}}-x_k)-
\exp(x_k-x_{k-1})+\exp(x_k-\wx_{k-1}).
\end{equation}

The same B\"acklund transformation as for the continuous time systems 
relates these systems of difference equations to the discrete time relativistic
Toda lattice:
\begin{equation}\label{dold}
\frac{\exp(\wx_k-x_k)-1}{\exp(x_k-\undertilde{x_k})-1}=
\frac{\Big(1+g^2\exp(x_{k+1}-x_k)\Big)}
{\Big(1+g^2\exp(\undertilde{x_{k+1}}-x_k)\Big)}
\frac{\Big(1+g^2\exp(x_k-\wx_{k-1})\Big)}
{\Big(1+g^2\exp(x_k-x_{k-1})\Big)}
\end{equation}

A modification of the construction leading to the discrete time systems above
allows to derive several further nice discretizations. For example, for
the lattice (\ref{new1}):
\begin{equation}\label{dmix1}
\exp(\wx_k-2x_k+\undertilde{x_k})=\frac
{\Big(1+h\exp(x_{k+1}-x_k)\Big)\Big(1-h\exp(\undertilde{x_{k+1}}-x_k)\Big)}
{\Big(1+h\exp(x_k-x_{k-1})\Big)\Big(1-h\exp(x_k-\wx_{k-1})\Big)},
\end{equation}
and for the relativistic Toda lattice (\ref{old}):
\begin{equation}\label{dmix2}
\frac{\exp(-\wx_k+x_k)-1}{\exp(-x_k+\undertilde{x_k})-1}=
\frac{\Big(1+g^2\exp(x_{k+1}-x_k)\Big)}
{\Big(1+g^2\exp(\undertilde{x_{k+1}}-x_k)\Big)}
\frac{\Big(1+g^2\exp(x_k-\wx_{k-1})\Big)}
{\Big(1+g^2\exp(x_k-x_{k-1})\Big)}.
\end{equation}
(The last system resembles very much the previous discretization of the 
relativistic Toda lattice (\ref{dold}), however the relation between them 
is far from trivial).

All the systems above (continuous and discrete time ones) may be considered
either on an infinite lattice ($k\in{\Bbb Z}$), or on a finite one 
($1\le k\le N$). In the last case one of the two types of boundary conditions 
may be imposed: open--end ($x_0=\infty$, $x_{N+1}=-\infty$) or periodic
($x_0\equiv x_N$, $x_{N+1}\equiv x_1$). We shall be concerned only with the
finite lattices here, consideration of the infinite ones being to a large
extent similar.

The last remark: the ''list'' of references at the end of this paper might 
look strange; in fact most of the references listed in \cite{DRTL} are relevant, 
and we do not reproduce them here only in order to save place. An interested 
reader is advised to consult these references.

\setcounter{equation}{0}
\section{Simplest flows of the relativistic Toda hierarchy\newline
and their bi--Hamiltonian structure}

We consider in this section two simplest flows of the relativistic Toda 
hierarchy. The first of them is:
\begin{equation}\label{RTL+}
\dot{d}_k=d_k(c_k-c_{k-1}), \quad
\dot{c}_k=c_k(d_{k+1}+c_{k+1}-d_k-c_{k-1}).
\end{equation}
The second flow of the relativistic Toda hierarchy is:
\begin{equation}\label{RTL-}
\dot{d}_k=d_k\left(\frac{c_k}{d_kd_{k+1}}-
\frac{c_{k-1}}{d_{k-1}d_k}\right), \quad
\dot{c}_k=c_k\left(\frac{1}{d_k}-\frac{1}{d_{k+1}}\right).
\end{equation}
They may be considered either under open--end boundary conditions
($d_{N+1}=c_0=c_N=0$), or under periodic ones (all the subscripts are
taken (mod $N$), so that $d_{N+1}\equiv d_1$, $c_0\equiv c_N$, 
$c_{N+1}\equiv c_1$). 

The both lattices (\ref{new1}), (\ref{old}) arise from the flow (\ref{RTL+}) 
under two different parametrizations of the variables $(c,d)$ by the 
canonically conjugated variables $(x,p)$. Analogously, the both lattices
(\ref{new2}), (\ref{old}) may be considered as arising on the same way 
from the flow (\ref{RTL-}).

We discuss now a Hamiltonian structure of the both flows (\ref{RTL+}), 
(\ref{RTL-}). It is easy to see that they are Hamiltonian with respect 
to two different compatible Poisson brackets. The first bracket is linear:
\begin{equation}\label{l br}
\{c_k,d_{k+1}\}_1=-c_k, \quad \{c_k,d_k\}_1=c_k,\quad \{d_k,d_{k+1}\}_1=c_k,
\end{equation}
(only the non--vanishing brackets are written down), and Hamiltonian functions
generating the flows (\ref{RTL+}), (\ref{RTL-}) in this bracket are equal to
\begin{equation}\label{H 1}
H^{(1)}_+=\frac{1}{2}\sum_{k=1}^N(d_k+c_{k-1})^2+\sum_{k=1}^N(d_k+c_{k-1})c_k,
\quad H^{(1)}_-=-\sum_{k=1}^N\log(d_k).
\end{equation}
The second Poisson bracket is quadratic:
\begin{equation}\label{q br}
\{c_k,c_{k+1}\}_2=-c_kc_{k+1}, \quad \{c_k,d_{k+1}\}_2=-c_kd_{k+1}, \quad
\{c_k,d_k\}_2=c_kd_k
\end{equation}
the corresponding Hamiltonian functions being
\begin{equation}\label{H 2}
H^{(2)}_+=\sum_{k=1}^N (d_k+c_k),\quad 
H^{(2)}_-=\sum_{k=1}^N\frac{d_k+c_k}{d_kd_{k+1}}.
\end{equation}

We turn now to the integrable discretizations of the flows (\ref{RTL+}),
(\ref{RTL-}), derived in \cite{DRTL}.

An integrable discretization of the flow (\ref{RTL+}) is given by the difference 
equations
\begin{equation}\label{dRTL+}
\widetilde{d}_k=d_k\frac{{\goth a}_{k+1}-hd_{k+1}}{{\goth a}_k-hd_k},\quad
\widetilde{c}_k=c_k\frac{{\goth a}_{k+1}+hc_{k+1}}{{\goth a}_k+hc_k}, 
\end{equation}
where ${\goth a}_k={\goth a}_k(c,d)$ is defined as a unique set of functions 
satisfying the recurrent relation
\begin{equation}\label{recur+}
{\goth a}_k=1+hd_k+\frac{hc_{k-1}}{{\goth a}_{k-1}}
\end{equation}
together with an asymptotic relation
\begin{equation}\label{as a}
{\goth a}_k=1+h(d_k+c_{k-1})+O(h^2).
\end{equation}
In the open--end case, due to $c_0=0$, we obtain from (\ref{recur+}) the 
following finite continued fractions expressions for ${\goth a}_k$:
\[
{\goth a}_1=1+hd_1;\quad 
{\goth a}_2=1+hd_2+\frac{hc_1}{1+hd_1};\quad\ldots\quad;
\]
\[
{\goth a}_N=1+hd_N+\frac{hc_{N-1}}{1+hd_{N-1}+
\displaystyle\frac{hc_{N-2}}{1+hd_{N-2}+
\parbox[t]{1.0cm}{$\begin{array}{c}\\  \ddots\end{array}$}
\parbox[t]{2.2cm}{$\begin{array}{c}
 \\  \\+\displaystyle\frac{hc_1}{1+hd_1}\end{array}$}}}.
\]
In the periodic case  (\ref{recur+}), (\ref{as a}) uniquely define 
${\goth a}_k$'s as $N$-periodic infinite continued fractions. It can be 
proved that for $h$ small enough these continued fractions converge and their 
values satisfy (\ref{as a}).

An integrable discretization of the flow (\ref{RTL-}) is given by the difference 
equations
\begin{equation}\label{dRTL-}
\widetilde{d}_k=
d_{k+1}\frac{d_k-h{\goth d}_{k-1}}{d_{k+1}-h{\goth d}_k}, \quad
\widetilde{c}_k=
c_{k+1}\frac{c_k+h{\goth d}_k}{c_{k+1}+h{\goth d}_{k+1}},
\end{equation}
where ${\goth d}_k={\goth d}_k(c,d)$ is defined as a unique set of functions 
satisfying the recurrent relation
\begin{equation}\label{recur-}
\frac{c_k}{{\goth d}_k}=d_k-h-h{\goth d}_{k-1},
\end{equation}
together with an asymptotic relation
\begin{equation}\label{as d}
{\goth d}_k=\frac{c_k}{d_k}+O(h).
\end{equation}
In the open--end case we obtain from (\ref{recur-}) the 
following finite continued fractions expressions for ${\goth d}_k$:
\[
{\goth d}_1=\frac{c_1}{d_1-h};\quad
{\goth d}_2=\frac{c_2}{d_2-h-\displaystyle\frac{hc_1}{d_1-h}};\quad
\ldots\quad;
\]
\[
{\goth d}_{N-1}=\frac{c_{N-1}}{d_{N-1}-h-\displaystyle
\frac{hc_{N-2}}{d_{N-2}-h-
\parbox[t]{1.0cm}{$\begin{array}{c}\\  \ddots\end{array}$}
\parbox[t]{2cm}{$\begin{array}{c}
 \\  \\-\displaystyle\frac{hc_1}{d_1-h}\end{array}$}}}.
\]
In the periodic case  (\ref{recur-}), (\ref{as d}) uniquely define 
${\goth d}_k$'s as $N$-periodic infinite continued fractions. It can be 
proved that for $h$ small enough these continued fractions converge and their 
values satisfy (\ref{as d}).

It can be proved \cite{DRTL} that the maps (\ref{dRTL+}), (\ref{dRTL-}) are
Poisson with respect to the both brackets (\ref{l br}), (\ref{q br}).

\setcounter{equation}{0}
\section{Lax representations}
Recall \cite{DRTL} that both continuous time systems (\ref{RTL+}), 
(\ref{RTL-}) and discrete time ones (\ref{dRTL+}), (\ref{dRTL-}) admit Lax 
representations, the Lax matrices being the same for the both cases.

Namely, the following statement holds.
Introduce two $N$ by $N$ matrices depending on the phase space coordinates 
$c_k, d_k$ and (in the periodic case) on the additional parameter $\lambda$:
\begin{eqnarray}
L(c,d,\lambda) & = & \sum_{k=1}^N d_kE_{kk}+\lambda\sum_{k=1}^N E_{k+1,k},\\
U(c,d,\lambda) & = & \sum_{k=1}^N E_{kk}-\lambda^{-1}\sum_{k=1}^N 
c_kE_{k,k+1}.
\end{eqnarray}
Here $E_{jk}$ stands for the matrix whose only nonzero entry on the intersection
of the $j$th row and the $k$th column is equal to 1. In the periodic case we
have $E_{N+1,N}=E_{1,N}, E_{N,N+1}=E_{N,1}$; in the open--end case we set
$\lambda=1$, and $E_{N+1,N}=E_{N,N+1}=0$. Consider also following two matrices:
\begin{equation}
T_+(c,d,\lambda)=L(c,d,\lambda)U^{-1}(c,d,\lambda),
\quad T_-(c,d,\lambda)=U^{-1}(c,d,\lambda)L(c,d,\lambda).
\end{equation}

{\bf Proposition 1.} {\it The flow} (\ref{RTL+}) {\it is equivalent to
the following matrix differential equations:
\begin{equation}
\dot{L}=LB-AL, \quad \dot{U}=UB-AU,
\end{equation}
which imply also
\begin{equation}
\dot{T}_+=\left[ T_+,A\right], \quad \dot{T}_-=\left[ T_-,B\right] ,
\end{equation}
where}
\begin{eqnarray}
A(c,d,\lambda) & = & \sum_{k=1}^N(d_k+c_{k-1})E_{kk}+\lambda
\sum_{k=1}^N E_{k+1,k},\\
B(c,d,\lambda) & = & \sum_{k=1}^N(d_k+c_k)E_{kk}+\lambda
\sum_{k=1}^N E_{k+1,k}.
\end{eqnarray}

{\bf Proposition 2.} {\it The map} (\ref{dRTL+}) {\it is equivalent to the 
following matrix difference equations:
\begin{equation}
\widetilde{L}={\rm\bf A}^{-1}L{\rm\bf B},\quad 
\widetilde{U}={\rm\bf A}^{-1}U{\rm\bf B},
\end{equation}
which imply also
\begin{equation}
\widetilde{T}_+={\rm\bf A}^{-1}T_+{\rm\bf A}, \quad
\widetilde{T}_-={\rm\bf B}^{-1}T_-{\rm\bf B},
\end{equation}
where
\begin{equation}\label{bA}
{\rm\bf A}(c,d,\lambda)=\sum_{k=1}^N{\goth a}_kE_{kk}
+h\lambda\sum_{k=1}^NE_{k+1,k},
\end{equation}
\begin{equation}\label{bB}
{\rm\bf B}(c,d,\lambda)=\sum_{k=1}^N{\goth b}_kE_{kk}
+h\lambda\sum_{k=1}^NE_{k+1,k},
\end{equation}
and the quantities ${\goth b}_k$ are defined by}
\begin{equation}\label{b}
{\goth b}_k={\goth a}_k\frac{{\goth a}_{k+1}-hd_{k+1}}{{\goth a}_k-hd_k}=
{\goth a}_{k-1}\frac{{\goth a}_k+hc_k}{{\goth a}_{k-1}+hc_{k-1}}.
\end{equation}

Note that the compatibility of the two expressions for ${\goth b}_k$ in
(\ref{b}) is an immediate consequence of (\ref{recur+}), and that it follows 
from (\ref{b}), (\ref{as a}) that
\begin{equation}\label{as b}
{\goth b}_k=1+h(d_k+c_k)+O(h^2).
\end{equation}

{\bf Proposition 3.} {\it The flow} (\ref{RTL-}) {\it is equivalent to
the following matrix differential equations:
\begin{equation}
\dot{L}=LD-CL, \quad \dot{U}=UD-CU,
\end{equation}
which imply also
\begin{equation}
\dot{T}_+=\left[ T_+,C\right], \quad \dot{T}_-=\left[ T_-,D\right] ,
\end{equation}
where}
\begin{eqnarray}
C(c,d) & = & -\lambda^{-1}\sum_{k=1}^{N} \frac{c_k}{d_{k+1}}E_{k,k+1}, \\
D(c,d) & = & -\lambda^{-1}\sum_{k=1}^{N} \frac{c_k}{d_k}E_{k,k+1}.
\end{eqnarray}

{\bf Proposition 4.} {\it The map} (\ref{dRTL-}) {\it is equivalent to the 
following matrix difference equations:
\begin{equation}
\widetilde{L}={\rm\bf C}L{\rm\bf D}^{-1},\quad 
\widetilde{U}={\rm\bf C}U{\rm\bf D}^{-1}
\end{equation}
which imply also
\begin{equation}
\widetilde{T}_+={\rm\bf C}T_+{\rm\bf C}^{-1}, \quad
\widetilde{T}_-={\rm\bf D}T_-{\rm\bf D}^{-1},
\end{equation}
where
\begin{equation}\label{bC}
{\rm\bf C}(c,d,\lambda)=
\sum_{k=1}^NE_{kk}+h\lambda^{-1}\sum_{k=1}^{N}{\goth c}_kE_{k,k+1},
\end{equation}
\begin{equation}\label{bD}
{\rm\bf D}(c,d,\lambda)=
\sum_{k=1}^NE_{kk}+h\lambda^{-1}\sum_{k=1}^{N}{\goth d}_kE_{k,k+1},
\end{equation}
and the quantities ${\goth c}_k$ are defined by}
\begin{equation}\label{c}
{\goth c}_k={\goth d}_k\frac{d_k-h{\goth d}_{k-1}}{d_{k+1}-h{\goth d}_k}
={\goth d}_{k+1}\frac{c_k+h{\goth d}_k}{c_{k+1}+h{\goth d}_{k+1}}.
\end{equation}

The compatibility of the two expressions for ${\goth c}_k$ in
(\ref{c}) is an immediate consequence of (\ref{recur-}), and it follows 
from (\ref{c}), (\ref{as d}) that
\begin{equation}\label{as c}
{\goth c}_k=\frac{c_k}{d_{k+1}}+O(h).
\end{equation}

The spectral invariants of the matrices $T_{\pm}(c,d,\lambda)$ serve as 
integrals of motion for the flows (\ref{RTL+}), (\ref{RTL-}), as well as for 
the maps (\ref{dRTL+}), (\ref{dRTL-}). 
In particular, it is easy to see that the Hamiltonian functions (\ref{H 1}),
(\ref{H 2}) are spectral invariants of the Lax matrices:
\[
H^{(1)}_+=\frac{1}{2}{\rm tr}(T_{\pm}^2),\quad H^{(1)}_-=-{\rm tr}\log(T_{\pm}),
\]
\[
H^{(2)}_+={\rm tr}(T_{\pm}),\quad H^{(2)}_-={\rm tr}(T_{\pm}^{-1}).
\]
Moreover, it can be proved \cite{DRTL} that the maps (\ref{dRTL+}), (\ref{dRTL-})
admit in both Poisson brackets (\ref{l br}), (\ref{q br}) interpolation by 
Hamiltonian flows with the Hamiltonian fuctions being certain spectral 
invariants of the matrices $T_{\pm}$.

\setcounter{equation}{0}
\section{Parametrization of the linear bracket\newline
by canonically conjugated variables}

In what follows we shall consider different Poisson maps from the standard 
symplectic space ${\Bbb R}^{2N}(x,p)$ into the Poisson space 
${\Bbb R}^{2N}(c,d)$, the latter being equipped with different Poisson brackets, 
while the former always with the canonical one:
\begin{equation}\label{can PB}
\{x_k,x_j\}=\{p_k,p_j\}=0,\quad \{p_k,x_j\}=\delta_{kj}.
\end{equation}
We shall call such maps {\it parametrizations} of the corresponding Poisson 
bracket on ${\Bbb R}^{2N}(c,d)$ through canonically conjugated variables
$(x,p)$.

For example, the linear Poisson bracket (\ref{l br}) may be parametrized 
by the canonically conjugated variables $(x,p)$ according to the formulas:
\begin{equation}\label{l par}
d_k=p_k-\exp(x_k-x_{k-1}),\quad c_k=\exp(x_{k+1}-x_k).
\end{equation}

Let us see how do the equations of motion look in this parametrization. We
start with (\ref{RTL+}), (\ref{dRTL+}).

Obviously, the function $H^{(1)}_+$ takes the form
\begin{equation}\label{H 1 + in xp}
H^{(1)}_+=\frac{1}{2}\sum_{k=1}^Np_k^2+\sum_{k=1}^Np_k\exp(x_{k+1}-x_k).
\end{equation}
Correspondingly, the flow (\ref{RTL+}) takes the form of canonical equations 
of motion:
\begin{eqnarray*}
\dot{x}_k & = & \partial H^{(1)}_+/\partial p_k=p_k+\exp(x_{k+1}-x_k),\\
\dot{p}_k & = & -\partial H^{(1)}_+/\partial x_k=
p_k\exp(x_{k+1}-x_k)-p_{k-1}\exp(x_k-x_{k-1}).
\end{eqnarray*}
As an immediate consequence of these equations one gets the Newtonian equations
of motion (\ref{new1}). A standard procedure allows to find a Lagrangian 
formulation of these equations. Indeed, one has to express
\begin{equation}\label{H to L}
{\cal L}=\sum_{k=1}^N \dot{x}_kp_k-H
\end{equation}
in terms of $(x_k,\dot{x}_k)$, which in the present case leads to
\begin{equation}\label{new1 Lagr}
{\cal L}^{(1)}_+(x,\dot{x})
=\frac{1}{2}\sum_{k=1}^N\Big(\dot{x}_k-\exp(x_{k+1}-x_k)\Big)^2.
\end{equation}
Note that the results of the previous section provide us with a Lax 
representation of our new lattice (\ref{new1}): one needs only to set in the 
formulas of the Proposition 1
\[
c_k=\exp(x_{k+1}-x_k),\quad d_k=\dot{x}_k-\exp(x_k-x_{k-1})-\exp(x_{k+1}-x_k).
\]

We turn now to a less straightforward case of discrete equations of motion.

{\bf Theorem 1.} {\it In the parametrization} (\ref{l par}) {\it the
equations of motion} (\ref{dRTL+}) {\it may be presented in the form of the 
following two equations:
\begin{equation}\label{dnew1:p}
hp_k = \exp(\wx_k-x_k)-\frac{1}{1-h\exp(x_k-\wx_{k-1})}+h\exp(x_k-x_{k-1})
-h\exp(x_{k+1}-x_k),
\end{equation}
\begin{equation}\label{dnew1:wp}
h\wip_k = \exp(\wx_k-x_k)-\frac{1}{1-h\exp(x_{k+1}-\wx_k)},
\end{equation}
which imply also the Newtonian equations of motion} (\ref{dnew1}).

{\bf Proof.} The second equation of motion in (\ref{dRTL+}) together with
the parametrization $c_k=\exp(x_{k+1}-x_k)$ implies that the following
quantity is constant, i.e. does not depend on $k$:
\[
\exp(-\wx_k+x_k)({\goth a}_k+hc_k)={\rm const}.
\]
Choosing this constant to be equal to 1, we get:
\begin{equation}\label{a+hc}
{\goth a}_k+hc_k=\exp(\wx_k-x_k),
\end{equation}
hence
\begin{equation}\label{a}
{\goth a}_k=\exp(\wx_k-x_k)-h\exp(x_{k+1}-x_k)
=\exp(\wx_k-x_k)\Big(1-h\exp(x_{k+1}-\wx_k)\Big).
\end{equation}

Substituting the last two formulas into (\ref{recur+}), we get
\begin{equation}\label{a-hd}
{\goth a}_k-hd_k=1+\frac{hc_{k-1}}{{\goth a}_{k-1}}
=\frac{1}{1-h\exp(x_k-\wx_{k-1})},
\end{equation}
or
\begin{equation}\label{dnew1:d}
hd_k=\exp(\wx_k-x_k)\Big(1-h\exp(x_{k+1}-\wx_k)\Big)
-\frac{1}{1-h\exp(x_k-\wx_{k-1})}.
\end{equation}
Now the first equation of motion in (\ref{dRTL+}) may be rewritten with the 
help of (\ref{a-hd}) as
\[
\widetilde{d}_k=d_k\frac{1-h\exp(x_k-\wx_{k-1})}{1-h\exp(x_{k+1}-\wx_k)},
\]
which together with (\ref{d}) implies:
\begin{equation}\label{dnew1:wd}
h\widetilde{d}_k=\exp(\wx_k-x_k)\Big(1-h\exp(x_k-\wx_{k-1})\Big)
-\frac{1}{1-h\exp(x_{k+1}-\wx_k)}.
\end{equation}
Under the parametrization $d_k=p_k-\exp(x_k-x_{k-1})$ the equations 
(\ref{dnew1:d}), (\ref{dnew1:wd}) are equivalent to (\ref{dnew1:p}), 
(\ref{dnew1:wp}). \qed

The Lax representations for the system (\ref{dnew1}) is given by the 
Proposition 2, where the expressions for the coefficients 
$c_k$, $d_k$, ${\goth a}_k$, ${\goth b}_k$ in terms of the variables $x_k$ and 
their discrete time updates $\wx_k$ are given by $c_k=\exp(x_{k+1}-x_k)$,
(\ref{dnew1:d}), (\ref{dnew1:wd}), (\ref{a}), and
\[
{\goth b}_k=\exp(\wx_k-x_k)\Big(1-h\exp(x_k-\wx_{k-1})\Big).
\]
(the last formula following from (\ref{b}), (\ref{a}), and (\ref{a-hd})).

Note also that the equations (\ref{dnew1:p}), (\ref{dnew1:wp}) not only 
immediately imply (\ref{dnew1}) from the introduction, but, moreover, allow to 
find a Lagrangian interpretation of this equation. Indeed, the general theory 
says that if the equations of motion are represented in the Lagrange form
\begin{equation}\label{dLagr}
\partial\Big(\Lambda(\wx,x)+\Lambda(x,\undertilde{x})\Big)/\partial x_k=0,
\end{equation}
then the momenta $p_k$ canonically conjugated to $x_k$ are given by
\begin{equation}\label{dLagr:p}
p_k=-\partial\Lambda(\wx,x)/\partial x_k,
\end{equation}
so that
\begin{equation}\label{dLagr:wp}
\widetilde{p}_k=\partial\Lambda(\wx,x)/\partial \widetilde{x}_k.
\end{equation}
Identifying (\ref{dnew1:p}), (\ref{dnew1:wp}) with (\ref{dLagr:p}), 
(\ref{dLagr:wp}), respectively, we see that the Lagrange function for the 
equation (\ref{dnew1}) can be chosen in the form
\begin{equation}\label{dnew1 Lagr}
\Lambda^{(1)}_+(\wx,x)=\sum_{k=1}^N\varphi(\wx_k-x_k)
-h^{-1}\sum_{k=1}^N\log\Big(1-h\exp(x_{k+1}-\wx_k)\Big)-
\sum_{k=1}^N\exp(x_{k+1}-x_k),
\end{equation}
where
\[
\varphi(\xi)=h^{-1}\Big(\exp(\xi)-1-\xi\Big).
\]
Obviously, this function serves a finite difference approximation to
(\ref{new1 Lagr}).

We turn now to the equations of motion (\ref{RTL-}), (\ref{dRTL-}), and find
out how do they look in the parametrization (\ref{l par}).

The function $H^{(1)}_-$ takes the form
\begin{equation}\label{H 1 - in xp}
H^{(1)}_-=-\sum_{k=1}^N\log\Big(p_k-\exp(x_k-x_{k-1})\Big).
\end{equation}
Correspondingly, the canonical equations 
of motion for the flow (\ref{RTL-}) takes the form:
\begin{eqnarray*}
\dot{x}_k & = & \partial H^{(1)}_-/\partial p_k=
-\frac{1}{p_k-\exp(x_k-x_{k-1})},\\
\dot{p}_k & = & -\partial H^{(1)}_-/\partial x_k=
\frac{\exp(x_{k+1}-x_k)}{p_{k+1}-\exp(x_{k+1}-x_k)}-
\frac{\exp(x_k-x_{k-1})}{p_k-\exp(x_k-x_{k-1})}.
\end{eqnarray*}
As a consequence of these equations, one gets:
\[
p_k=\exp(x_k-x_{k-1})-\frac{1}{\dot{x}_k},\quad
\dot{p}_k=-\dot{x}_{k+1}\exp(x_{k+1}-x_k)+\dot{x}_k\exp(x_k-x_{k-1}),
\]
and the Newtonian equations of motion (\ref{new2}) follow. 
A standard procedure (\ref{H to L}) leads to a Lagrangian 
formulation of these equations. One has:
\begin{equation}\label{new2 Lagr}
{\cal L}^{(1)}_-(x,\dot{x})
=-\sum_{k=1}^N\log(\dot{x}_k)+\sum_{k=1}^N\dot{x}_k\exp(x_k-x_{k-1}).
\end{equation}
In order to get a Lax representation of the lattice (\ref{new2}) one needs only 
to set in the formulas of the Proposition 3:
\[
c_k=\exp(x_{k+1}-x_k),\quad d_k=-\frac{1}{\dot{x}_k}.
\]

Turn to the discrete equations of motion (\ref{dRTL-}), we get:

{\bf Theorem 2.} {\it In the parametrization} (\ref{l par}) {\it the
equations of motion} (\ref{dRTL-}) {\it may be presented in the form of the 
following two equations:
\begin{equation}\label{dnew2:p}
p_k = -\frac{h}{\exp(\wx_k-x_k)-1}+\exp(x_k-\wx_{k-1}),
\end{equation}
\begin{equation}\label{dnew2:wp}
\wip_k = -\frac{h}{\exp(\wx_k-x_k)-1}+\exp(x_{k+1}-\wx_k)
-\exp(\wx_{k+1}-\wx_k)+\exp(\wx_k-\wx_{k-1}),
\end{equation}
which imply also the Newtonian equations of motion} (\ref{dnew1}).

{\bf Proof.} The second equation of motion in (\ref{dRTL-}), rewritten as
\[
\wc_k=c_k\frac{1+h{\goth d}_k/c_k}{1+h{\goth d}_{k+1}/c_{k+1}},
\]
together with the parametrization $c_k=\exp(x_{k+1}-x_k)$ implies that the 
following quantity is constant, i.e. does not depend on $k$:
\[
\exp(\wx_k-x_k)\left(1+\frac{h{\goth d}_k}{c_k}\right)={\rm const}.
\]
Choosing this constant to be equal to 1, we get:
\begin{equation}\label{c/d+h}
\frac{c_k}{{\goth d}_k}+h=\frac{h}{\exp(\wx_k-x_k)-1},
\end{equation}
hence
\begin{equation}\label{d}
h{\goth d}_k=\exp(x_{k+1}-\wx_k)-\exp(x_{k+1}-x_k)=
-\exp(x_{k+1}-\wx_k)\Big(\exp(\wx_k-x_k)-1\Big).
\end{equation}
The recurrent relation (\ref{recur-}) implies:
\begin{equation}\label{d-hd}
d_k-h{\goth d}_{k-1}=\frac{c_k}{{\goth d}_k}+h,
\end{equation}
which together with (\ref{d}), (\ref{c/d+h}) implies:
\begin{equation}\label{dnew2:d}
d_k=-\frac{h}{\exp(\wx_k-x_k)-1}-
\exp(x_k-\wx_{k-1})\Big(\exp(\wx_{k-1}-x_{k-1})-1\Big).
\end{equation}
Now we can rewrite the first equation of motion in (\ref{dRTL-}), taking
into account (\ref{d-hd}), (\ref{c/d+h}):
\[
\widetilde{d}_k=d_{k+1}\;\frac{\exp(\wx_{k+1}-x_{k+1})-1}{\exp(\wx_k-x_k)-1}.
\]
The last equation together with (\ref{dnew2:d}) implies:
\begin{equation}\label{dnew2:wd}
\widetilde{d}_k=-\frac{h}{\exp(\wx_k-x_k)-1}-
\exp(x_{k+1}-\wx_k)\Big(\exp(\wx_{k+1}-x_{k+1})-1\Big).
\end{equation}
Under the parametrization $d_k=p_k-\exp(x_k-x_{k-1})$ the equations 
(\ref{dnew2:d}), (\ref{dnew2:wd}) are equivalent to (\ref{dnew2:p}), 
(\ref{dnew2:wp}). \qed

The Lax representations for the system (\ref{dnew2}) is given by the 
Proposition 4, where the expressions for the coefficients 
$c_k$, $d_k$, ${\goth d}_k$, ${\goth c}_k$ in terms of the variables $x_k$ and 
their discrete time updates $\wx_k$ are given by $c_k=\exp(x_{k+1}-x_k)$,
(\ref{dnew2:d}), (\ref{dnew2:wd}), (\ref{d}), and
\[
h{\goth c}_k=-\exp(x_{k+1}-\wx_k)\Big(\exp(\wx_{k+1}-x_{k+1})-1\Big)
\]
(the last formula following from (\ref{d}), (\ref{c}), and (\ref{c/d+h})).

Identifying (\ref{dnew2:p}), (\ref{dnew2:wp}) with (\ref{dLagr:p}), 
(\ref{dLagr:wp}), respectively, we get the Lagrange function for the 
equation (\ref{dnew2}) in the form
\[
\Lambda^{(1)}_-(\wx,x)=-h\sum_{k=1}^N\psi(\wx_k-x_k)
+\sum_{k=1}^N\Big[\exp(\wx_k-\wx_{k-1})-\exp(x_k-\wx_{k-1})\Big],
\]
where
\[
\psi(\xi)=\int_0^{\xi}\frac{d\eta}{\exp(\eta)-1}=\log(\exp(\xi)-1)-\xi.
\]
This function clearly is a finite difference approximation of (\ref{new2 Lagr}).

\setcounter{equation}{0}
\section{Parametrization of the quadratic bracket\newline
by canonically conjugated variables}

We give for completeness the results corresponding to another parametrization
of the variables $c_k, d_k$ by means of canonically conjugated variables 
$x_k,p_k$, namely the parametrization leading to the quadratic bracket
(\ref{q br}). The relativistic Toda lattice arises on this way. The 
corresponding formulas were given in \cite{DRTL}, but in an {\it ad hoc} manner,
without derivation. We take an opportunity to fill in this gap here.

The parametrization leding to the quadratic bracket (\ref{q br}) reads:
\begin{equation}\label{q par}
d_k=\exp(p_k), \quad c_k=g^2\exp(x_{k+1}-x_k+p_k),
\end{equation}
($g^2\in{\Bbb R}$ is a coupling constant). 

In terms of these variables
\begin{eqnarray}
H^{(2)}_+&=&\sum_{k=1}^N \exp(p_k)\Big(1+g^2\exp(x_{k+1}-x_k)\Big),\nonumber\\
H^{(2)}_-&=&\sum_{k=1}^N \exp(-p_k)\Big(1+g^2\exp(x_k-x_{k-1})\Big).
\label{H 2 in xp}
\end{eqnarray}

Hence the equation of motion corresponding to (\ref{RTL+}) take the canonical
form
\begin{eqnarray*}
\dot{x}_k &=&\partial H^{(2)}_+/\partial p_k=
\exp(p_k)\Big(1+g^2\exp(x_{k+1}-x_k)\Big),\\
\dot{p}_k &=&-\partial H^{(2)}_+/\partial x_k=
g^2\exp(x_{k+1}-x_k+p_k)-g^2\exp(x_k-x_{k-1}+p_{k-1}).
\end{eqnarray*}
This can be put into a Newtonian form (\ref{old}).

A standard procedure (\ref{H to L}) allows also to find a Lagrangian 
formulation of these equations, with a Lagrange function
\begin{equation}\label{Lagr+}
{\cal L}^{(2)}_+(x,\dot{x})=\sum_{k=1}^N[\dot{x}_k\log(\dot{x}_k)-\dot{x}_k]-
\sum_{k=1}^N\dot{x}_k\log\Big(1+g^2\exp(x_{k+1}-x_k)\Big).
\end{equation}

The Lax representation for these equations are given by the Proposition 1 with
\[
d_k=\dot{x}_k/\Big(1+g^2\exp(x_{k+1}-x_k)\Big),\quad 
c_k=g^2\exp(x_{k+1}-x_k)d_k.
\]

Analogously, the canonical equations of motion corresponding to (\ref{RTL-}) are:
\begin{eqnarray*}
\dot{x}_k &=&\partial H^{(2)}_-/\partial p_k=
-\exp(-p_k)\Big(1+g^2\exp(x_k-x_{k-1})\Big),\\
\dot{p}_k &=&-\partial H^{(2)}_-/\partial x_k=
g^2\exp(x_{k+1}-x_k-p_{k+1})-g^2\exp(x_k-x_{k-1}-p_k).
\end{eqnarray*}
The Newtonian equations following from these ones are just the same (\ref{old})
as before. they correspond, however, to a different form of a Lagrange function:
\begin{equation}\label{Lagr-}
{\cal L}^{(2)}_-(x,\dot{x})=-\sum_{k=1}^N[\dot{x}_k\log(-\dot{x}_k)-\dot{x}_k]+
\sum_{k=1}^N\dot{x}_k\log\Big(1+g^2\exp(x_k-x_{k-1})\Big).
\end{equation}
Respectively, the Lax representation for these equations is given by the
Proposition 3 with the identifications
\[
d_k=-\frac{1+g^2\exp(x_k-x_{k-1})}{\dot{x}_k},\quad
c_k=g^2\exp(x_{k+1}-x_k)d_k.
\]

We turn now to the discrete time systems (\ref{dRTL+}), (\ref{dRTL-}).

{\bf Theorem 3.} {\it In the parametrization} (\ref{q par}) {\it the map} 
(\ref{dRTL+}) {\it takes the form of the following two equations:
\begin{eqnarray}
h\exp(p_k) & = & \frac{\Big(\exp(\wx_k-x_k)-1\Big)}
{\Big(1+g^2\exp(x_{k+1}-x_k)\Big)}\:
\frac{\Big(1+g^2\exp(x_k-x_{k-1})\Big)}{\Big(1+g^2\exp(x_k-\wx_{k-1})\Big)},
\label{dold1:p}\\
h\exp(\widetilde{p}_k) & = &
\frac{\Big(\exp(\wx_k-x_k)-1\Big)}{\Big(1+g^2\exp(x_{k+1}-\wx_k)\Big)}.
\label{dold1:wp}
\end{eqnarray}
This implies also a Newtonian form} (\ref{dold}) {\it of equations of motion.}

{\bf Proof.}  From  (\ref{dRTL+}) it follows:
\[
\frac{\widetilde{c}_k}{\widetilde{d}_k}=\frac{c_k}{d_k}\;
\frac{{\goth a}_{k+1}+hc_{k+1}}{{\goth a}_{k+1}-hd_{k+1}}\;
\frac{{\goth a}_k-hd_k}{{\goth a}_k+hc_k}.
\]
Since $c_k/d_k=g^2\exp(x_{k+1}-x_k)$, this implies that the following quantity
is constant, i.e. does not depend on $k$:
\[
\exp(\wx_k-x_k)\frac{{\goth a}_k-hd_k}{{\goth a}_k+hc_k}={\rm const}.
\]
Setting this constant equal to 1, we get:
\[
\frac{{\goth a}_k+hc_k}{{\goth a}_k-hd_k}=\exp(\wx_k-x_k).
\]
This implies:
\begin{eqnarray}
{\goth a}_k &=&h\frac{d_k\exp(\wx_k-x_k)+c_k}{\exp(\wx_k-x_k)-1}\nonumber\\
&=&hd_k
\frac{\exp(\wx_k-x_k)\Big(1+g^2\exp(x_{k+1}-\wx_k)\Big)}{\exp(\wx_k-x_k)-1}.
\label{aux}
\end{eqnarray}
As a consequence, we get:
\begin{eqnarray}
{\goth a}_k+hc_k &=&h\exp(\wx_k-x_k)\frac{d_k+c_k}{\exp(\wx_k-x_k)-1}\nonumber\\
&=&hd_k
\frac{\exp(\wx_k-x_k)\Big(1+g^2\exp(x_{k+1}-x_k)\Big)}{\exp(\wx_k-x_k)-1}.
\label{a+hc 2}
\end{eqnarray}
Substituting (\ref{aux}), (\ref{a+hc 2}) in the recurrent relation (\ref{recur+}),
we get:
\begin{equation}\label{a-hd 2}
{\goth a}_k-hd_k=1+\frac{hc_{k-1}}{{\goth a}_{k-1}}
=\frac{1+g^2\exp(x_k-x_{k-1})}{1+g^2\exp(x_k-\wx_{k-1})}.
\end{equation}
Substituting in the left--hand side of this formula the expression (\ref{aux})
for ${\goth a}_k$, we arrive at:
\begin{equation}\label{dold1:d}
hd_k=\frac{\Big(\exp(\wx_k-x_k)-1\Big)}{\Big(1+g^2\exp(x_{k+1}-x_k)\Big)}\:
\frac{\Big(1+g^2\exp(x_k-x_{k-1})\Big)}{\Big(1+g^2\exp(x_k-\wx_{k-1})\Big)},
\end{equation}
Further, from the first equation of motion in (\ref{dRTL+}) and (\ref{a-hd 2})
it follows:
\begin{equation}\label{dold1:wd}
h\widetilde{d}_k =
\frac{\exp(\wx_k-x_k)-1}{1+g^2\exp(x_{k+1}-\wx_k)}.
\end{equation}
Now (\ref{dold1:p}), (\ref{dold1:wp}) follow from (\ref{dold1:d}),
(\ref{dold1:wd}) under the parametrization $d_k=\exp(p_k)$. \qed

The Lax representation for the system (\ref{dold1:p}), (\ref{dold1:wp}) is
given by the Proposition 2 with the following expressions through $x_k$, 
$\wx_k$: (\ref{dold1:d}), (\ref{dold1:wd}) for $d_k$, $\widetilde{d}_k$;
$c_k=g^2\exp(x_{k+1}-x_k)d_k$;
\begin{eqnarray}
{\goth a}_k & = & \exp(\wx_k-x_k)
\frac{\Big(1+g^2\exp(x_{k+1}-\wx_k)\Big)}{\Big(1+g^2\exp(x_{k+1}-x_k)\Big)}\:
\frac{\Big(1+g^2\exp(x_k-x_{k-1})\Big)}{\Big(1+g^2\exp(x_k-\wx_{k-1})\Big)},
\label{a in x}\\
{\goth b}_k & = & \exp(\wx_k-x_k).\label{b in x}
\end{eqnarray}
Indeed, (\ref{a in x}) follows from (\ref{aux}), (\ref{dold1:d}), and 
(\ref{b in x}) follows from (\ref{b}), (\ref{a in x}), and (\ref{a-hd 2}).

Identifying (\ref{dold1:p}) and (\ref{dold1:wp}) with (\ref{dLagr:p}), 
(\ref{dLagr:wp}), we get a Lagrange function 
\begin{equation}\label{dRTL+ Lagr}
\Lambda^{(2)}_+(\wx,x)=\sum_{k=1}^N\Phi(\wx_k-x_k)+\sum_{k=1}^{N}\left[
\Psi(x_{k+1}-\wx_k)-\Psi(x_{k+1}-x_k)\right],
\end{equation}
where the two functions $\Phi(\xi), \Psi(\xi)$ are defined by
\begin{equation}\label{PP}
\Phi(\xi)=\int_0^{\xi}\log\left|\frac{\exp(\eta)-1}{h}\right|d\eta,\quad
\Psi(\xi)=\int_0^{\xi}\log\left(1+g^2\exp(\eta)\right)d\eta.
\end{equation}
It is easy to see that this Lagrangian function serves as a finite difference
approximation to (\ref{Lagr+}).

{\bf Theorem 4.} {\it In the parametrization} (\ref{q par}) {\it the map} 
(\ref{dRTL-}) {\it takes the form of the following two equations: 
\begin{eqnarray}
\exp(p_k) & = & \frac{h\Big(1+g^2\exp(x_k-\wx_{k-1})\Big)}
{\Big(1-\exp(\wx_k-x_k)\Big)},
\label{dold2:p}\\
\exp(\widetilde{p}_k) & = & 
\frac{h\Big(1+g^2\exp(\wx_k-\wx_{k-1})\Big)}{\Big(1-\exp(\wx_k-x_k)\Big)}\:
\frac{\Big(1+g^2\exp(x_{k+1}-\wx_k)\Big)}{\Big(1+g^2\exp(\wx_{k+1}-\wx_k)\Big)}.
\label{dold2:wp}
\end{eqnarray}
This implies the same Newtonian equations of motion} (\ref{dold})
{\it as for the map} (\ref{dRTL+}).

{\bf Proof.} From (\ref{dRTL-}), (\ref{recur-}) we deduce:
\[
\frac{\widetilde{c}_k}{\widetilde{d}_k}=
\frac{c_{k+1}}{d_{k+1}}\;\frac{c_k+h{\goth d}_k}{d_k-h{\goth d}_{k-1}}\;
\frac{d_{k+1}-h{\goth d}_k}{c_{k+1}+h{\goth d}_{k+1}}=
\frac{c_{k+1}}{d_{k+1}}\,\frac{{\goth d}_k}{{\goth d}_{k+1}}.
\]
Because of $c_k/d_k=g^2\exp(x_{k+1}-x_k)$, this implies that the following 
quantity does not depend on $k$:
\[
{\goth d}_k\exp(\wx_k-x_{k+1})={\rm const}.
\]
Setting this constant equal to $g^2$, we get:
\begin{equation}\label{d 2}
{\goth d}_k = g^2\exp(x_{k+1}-\wx_k).
\end{equation}
Substituting this formula into the recurrence (\ref{recur-}), we get:
\begin{equation}\label{dold2:d}
d_k = \frac{h\Big(1+g^2\exp(x_k-\wx_{k-1})\Big)}{\Big(1-\exp(\wx_k-x_k)\Big)}.
\end{equation}
Formulas 
(\ref{dold2:p}), (\ref{dold2:d}) imply also
\begin{equation}\label{d-hd 2}
d_k-h{\goth d}_{k-1}=\frac{h\Big(1+g^2\exp(\wx_k-\wx_{k-1})\Big)}
{\Big(1-\exp(\wx_k-x_k)\Big)}.
\end{equation}
This last formula together with the first equation in (\ref{dRTL-}) 
implies:
\begin{equation}\label{dold2:wd}
\widetilde{d}_k = 
\frac{h\Big(1+g^2\exp(\wx_k-\wx_{k-1})\Big)}{\Big(1-\exp(\wx_k-x_k)\Big)}\:
\frac{\Big(1+g^2\exp(x_{k+1}-\wx_k)\Big)}{\Big(1+g^2\exp(\wx_{k+1}-\wx_k)\Big)}.
\end{equation}
Finally, (\ref{dold2:p}), (\ref{dold2:wp}) is equivalent to (\ref{dold2:d}), 
(\ref{dold2:wd}), due to the parametrization $d_k=\exp(p_k)$. \qed

The Lax representation for the system (\ref{dold2:p}), (\ref{dold2:wp}) is
given by the Proposition 4 with the following expressions through $x_k$, 
$\wx_k$: (\ref{dold2:d}), (\ref{dold2:wd}) for $d_k$, $\widetilde{d}_k$;
$c_k=g^2\exp(x_{k+1}-x_k)d_k$; (\ref{d 2}) for ${\goth d}_k$; and
\[
{\goth c}_k = g^2\exp(x_{k+1}-\wx_k)
\frac{\Big(1-\exp(\wx_{k+1}-x_{k+1})\Big)}{\Big(1-\exp(\wx_k-x_k)\Big)}\:
\frac{\Big(1+g^2\exp(\wx_k-\wx_{k-1})\Big)}{\Big(1+g^2\exp(\wx_{k+1}-\wx_k)\Big)}.
\]
The last formula follows from (\ref{c}), (\ref{d 2}), and (\ref{d-hd 2}).

Identifying (\ref{dold2:p}) and (\ref{dold2:wp}) with (\ref{dLagr:p}), 
(\ref{dLagr:wp}), we get a Lagrange function 
\begin{equation}\label{dRTL- Lagr}
\Lambda^{(2)}_-(\wx,x)=-\sum_{k=1}^N\Phi(\wx_k-x_k)+\sum_{k=1}^{N}\left[
\Psi(\wx_k-\wx_{k-1})-\Psi(x_k-\wx_{k-1})\right],
\end{equation}
wit the same functions $\Phi(\xi), \Psi(\xi)$ (\ref{PP}) as before.
It is easy to see that this Lagrangian function is a finite difference
approximation to (\ref{Lagr-}).

\setcounter{equation}{0}
\section{Parametrization of the mixed brackets\newline
by canonically conjugated variables}

It turns out that there exist still another parametrizations of the variables
$(c,d)$ by canonically conjugated variables $(x,p)$ leading to interesting
discretizations. As we shall see, these parametrizations lead to Poisson
brackets which are linear combinations of the two homogeneous ones (\ref{l br})
and (\ref{q br}). In some sense (which will be clear from the proofs of the 
Theorems below) these parametrizations are specially designed to obtain nice
Newtonian equations from the maps (\ref{dRTL+}), (\ref{dRTL-}).

We start from the parametrization leading the linear combination 
$\{\cdot,\cdot\}_1+h\{\cdot,\cdot\}_2$, which turns out to admit nice 
discrete Newtonian formulation when applied to (\ref{dRTL+}).
Clearly, we get an alternative discretization of the lattice (\ref{new1})
on this way. Consider following parametrization:
\begin{equation}\label{mix1 par}
hd_k=\exp(hp_k)-1-h\exp(x_k-x_{k-1}),\qquad 
c_k=\exp(x_{k+1}-x_k+hp_k).
\end{equation}
(Obviously, in the limit $h\to 0$ we recover the parametrization of the linear 
bracket (\ref{l par})). Simple calculations show that the Poisson brackets
between $(c,d)$ induced by (\ref{mix1 par}) read:
\[
\begin{array}{c}
\{c_{k+1},c_k\}=hc_{k+1}c_k,\quad \{d_{k+1},d_k\}=-c_k,\\ \\
\{d_{k+1},c_k\}=c_k+hd_{k+1}c_k,\quad \{d_k,c_k\}=-c_k-hd_kc_k,
\end{array}
\]
which is exactly $\{\cdot,\cdot\}_1+h\{\cdot,\cdot\}_2$. Let us look at the 
equations of motion generated by these $h$--dependent Poisson bracket.

{\bf Theorem 5.} {\it In the parametrization} (\ref{mix1 par}) {\it the map} 
(\ref{dRTL+}) {\it takes the form of the following two equations: 
\begin{eqnarray}
\exp(hp_k) & = & \exp(\wx_k-x_k)\;
\frac{\Big(1+h\exp(x_k-x_{k-1})\Big)\Big(1-h\exp(x_k-\wx_{k-1})\Big)}
{\Big(1+h\exp(x_{k+1}-x_k)\Big)},
\label{dmix1:p}\\
\exp(h\wip_k) & = & \exp(\wx_k-x_k)\Big(1-h\exp(x_{k+1}-\wx_k)\Big).
\label{dmix1:wp}
\end{eqnarray}
This implies the Newtonian equations of motion} (\ref{dmix1}).

{\bf Proof.} The crucial observation consists in the following: we can extract
from the relations (\ref{mix1 par}) the following consequence:
\[
\exp(hp_k)=1+hd_k+\frac{hc_{k-1}}{\exp(hp_{k-1})}.
\]
Comparing this with (\ref{recur+}), we see that ${\goth a}_k$ and 
$\exp(hp_k)$ satisfy one and the same recurrent relation. Due to the
uniqueness of its solution, they must coincide, so that we obtain:
\begin{equation}\label{a 3}
{\goth a}_k=\exp(hp_k).
\end{equation}
As a consequence, we get immediately:
\begin{equation}\label{a-hd 3}
{\goth a}_k-hd_k=1+h\exp(x_k-x_{k-1}),\qquad 
{\goth a}_k+hc_k=\exp(hp_k)\Big(1+h\exp(x_{k+1}-x_k)\Big).
\end{equation}
These expressions together with (\ref{mix1 par}), being substituted into 
(\ref{dRTL+}), allow to rewrite the latter in the form:
\begin{eqnarray}
\exp(h\wip_k)-h\exp(\wx_k-\wx_{k-1})&=&\exp(hp_k)\,
\frac{1+h\exp(x_{k+1}-x_k)}{1+h\exp(x_k-x_{k-1})}-h\exp(x_{k+1}-x_k);
\nonumber\\
\exp(\wx_{k+1}-\wx_k+h\wip_k)&=&\exp(x_{k+1}-x_k+hp_{k+1})\;
\frac{1+h\exp(x_{k+2}-x_{k+1})}{1+h\exp(x_{k+1}-x_k)}.\label{aux 3}
\end{eqnarray}
The formula arising when excluding $\exp(\wip_k)$ from these two equations,
can be we written after some manipulations as:
\[
\exp(x_{k+1}-\wx_{k+1}+hp_{k+1})\,
\frac{1+h\exp(x_{k+2}-x_{k+1})}{1+h\exp(x_{k+1}-x_k)}+h\exp(x_{k+1}-\wx_k)=
\]
\[
\exp(x_k-\wx_k+p_k)\,\frac{1+h\exp(x_{k+1}-x_k)}{1+h\exp(x_k-x_{k-1})}
+h\exp(x_k-\wx_{k-1}).
\]
So, the expression on the right--hand side is constant, i.e. does not depend 
on $k$. Setting this constant equal to 1, we arrive at (\ref{dmix1:p}).
Substituting (\ref{dmix1:p}) into (\ref{aux 3}), we get (\ref{dmix1:wp}). \qed

It is not difficult to extract from this proof the expressions for coefficients
of the matrices forming the Lax representation of the system (\ref{dmix1})
following from the Proposition 2. Also a Lagrangian formulation of this system
can be obtained in a standard way: the Lagrange function corresponding to
(\ref{dmix1:p}), (\ref{dmix1:wp}) is 
\begin{equation}
\Lambda_+^{({\rm mixed})}(\wx,x)=\sum_{k=1}^N\frac{(\wx_k-x_k)^2}{2h}
-h^{-1}\sum_{k=1}^N\Big[\phi_1(x_{k+1}-\wx_k)+\phi_2(x_{k+1}-x_k)\Big],
\end{equation}
where
\[
\phi_1(\xi)=\int_0^{\xi}\log\Big(1-h\exp(\eta)\Big)d\eta,\qquad
\phi_2(\xi)=\int_0^{\xi}\log\Big(1+h\exp(\eta)\Big)d\eta.
\]
This is a finite difference approximation to (\ref{new1 Lagr}) different
from (\ref{dnew1 Lagr}).

We turn now to another parametrization leading to a mixed Poisson bracket,
namely to the bracket $\{\cdot,\cdot\}_2-h\{\cdot,\cdot\}_1$.
The corresponding formulas are:
\begin{equation}\label{new par}
d_k=\exp(p_k)+h\Big(1+g^2\exp(x_k-x_{k-1})\Big),\qquad 
c_k=g^2\exp(x_{k+1}-x_k+p_k).
\end{equation}
As is easy to calculate, the resulting Poisson brackets between the variables 
$(c,d)$ are:
\[
\begin{array}{c}
\{c_{k+1},c_k\}=c_{k+1}c_k,\quad \{d_{k+1},d_k\}=-hc_k,\\ \\
\{d_{k+1},c_k\}=d_{k+1}c_k-hc_k,\quad \{d_k,c_k\}=-d_kc_k+hc_k.
\end{array}
\]
i.e. the linear combination $\{\cdot,\cdot\}_2-h\{\cdot,\cdot\}_1$. The 
equations arising from (\ref{dRTL-}) under this parametrization, naturally
approximate the relativistic Toda lattice (\ref{old}). 

{\bf Theorem 6.} {\it In the parametrization} (\ref{new par}) {\it the map} 
(\ref{dRTL-}) {\it takes the form of the following two equations: 
\begin{eqnarray}
\exp(p_k) & = & \frac{h\Big(1+g^2\exp(x_k-\wx_{k-1})\Big)}
{\Big(\exp(-\wx_k+x_k)-1\Big)},
\label{dmix2:p}\\
\exp(\widetilde{p}_k) & = & 
\frac{h\Big(1+g^2\exp(\wx_k-\wx_{k-1})\Big)}{\Big(\exp(-\wx_k+x_k)-1\Big)}\:
\frac{\Big(1+g^2\exp(x_{k+1}-\wx_k)\Big)}{\Big(1+g^2\exp(\wx_{k+1}-\wx_k)\Big)}.
\label{dmix2:wp}
\end{eqnarray}
This implies the Newtonian equations of motion} (\ref{dmix2}).

{\bf Proof.} This time the crucial observation consists in the following: as a
consequence of the relations (\ref{new par}) we have:
\[
\frac{c_k}{g^2\exp(x_{k+1}-x_k)}=d_k-h-hg^2\exp(x_k-x_{k-1}).
\]
Comparing this with (\ref{recur-}), we see that ${\goth d}_k$ and 
$g^2\exp(x_{k+1}-x_k)$ satisfy one and the same recurrent relation. Due to the
uniqueness of its solution, they must coincide, so that we obtain:
\begin{equation}\label{d 3}
{\goth d}_k=g^2\exp(x_{k+1}-x_k).
\end{equation}
As a consequence, we get immediately:
\begin{equation}\label{d-hd 3}
d_k-h{\goth d}_{k-1}=\exp(p_k)+h,\qquad 
c_k+h{\goth d}_k=g^2\exp(x_{k+1}-x_k)\Big(\exp(p_k)+h\Big).
\end{equation}
Substituting these expressions together with (\ref{new par}) into (\ref{dRTL-}), 
we bring the latter to the form:
\begin{eqnarray}
\exp(\wip_k)+hg^2\exp(\wx_k-\wx_{k-1})&=&\exp(p_k)+hg^2\exp(x_{k+1}-x_k)\;
\frac{\exp(p_k)+h}{\exp(p_{k+1})+h};\nonumber\\
\exp(\wx_{k+1}-\wx_k+\wip_k)&=&\exp(x_{k+1}-x_k+p_{k+1})\;
\frac{\exp(p_k)+h}{\exp(p_{k+1})+h}.\label{aux 4}
\end{eqnarray}
Excluding $\exp(\wip_k)$ from these two equations, we get the formula which
after some manipulations can be written as:
\[
\frac{\exp(x_{k+1}-\wx_{k+1}+p_{k+1})-hg^2\exp(x_{k+1}-\wx_k)}{\exp(p_{k+1})+h}=
\frac{\exp(x_k-\wx_k+p_k)-hg^2\exp(x_k-\wx_{k1})}{\exp(p_k)+h}.
\]
So, the expression on the right--hand side is constant, i.e. does not depend 
on $k$. Setting this constant equal to 1, we arrive at (\ref{dmix2:p}), which
is equivalent also to
\begin{equation}\label{ep+h}
\exp(p_k)+h=\frac{h\exp(-\wx_k+x_k)\Big(1+g^2\exp(\wx_k-\wx_{k-1})\Big)}
{\exp(-\wx_k+x_k)-1}.
\end{equation}
Finally, substituting (\ref{ep+h}) into (\ref{aux 4}), we get 
\[
\exp(\wip_k)=\exp(p_{k+1})\,\frac{\Big(\exp(-\wx_{k+1}+x_{k+1})-1\Big)}
{\Big(\exp(-\wx_k+x_k)-1\Big)}\,
\frac{\Big(1+g^2\exp(\wx_k-\wx_{k-1})\Big)}
{\Big(1+g^2\exp(\wx_{k+1}-\wx_k)\Big)}
\]
which together with (\ref{dmix2:p}) implies (\ref{dmix2:wp}). \qed

The Lax representation for (\ref{dmix2}) is given by the Proposition 4;
it is not difficult to extract from the proof above the expressions for
the entries of the matrices forming the Lax representation. Also the
Lagrangian formulation of (\ref{dmix2}) easily follows from (\ref{dmix2:p}),
(\ref{dmix2:wp}). The corresponding Lagrange function is
\begin{equation}\label{dmix2 Lagr}
\Lambda^{({\rm mixed})}_-(\wx,x)=\sum_{k=1}^N\Phi(-\wx_k+x_k)+
\sum_{k=1}^{N}\left[\Psi(\wx_k-\wx_{k-1})-\Psi(x_k-\wx_{k-1})\right],
\end{equation}
wit the functions $\Phi(\xi), \Psi(\xi)$ given in (\ref{PP}).

Let us note that, though the equations (\ref{dold}), (\ref{dmix2}) are very
similar, there exist no obvious changes of variables bringing one of them into
another. The only way to do this, i.e. to connect two corresponding sets of 
$(x_k,\wx_k)$, is to identify the corresponding $(c_k,d_k)$, given for
(\ref{dold}) by (\ref{q par}), (\ref{dold2:p}), (\ref{dold2:wp}), and for
(\ref{dmix2}) by (\ref{new par}), (\ref{dmix2:p}), (\ref{dmix2:wp}). The
resulting change of variables is a rather nontrivial B\"acklund transformation.

\setcounter{equation}{0}
\section{Conclusion}

The main message of the present note is following: the field of integrable 
systems of classical mechanics, even in its best studied parts, is far
from being exhausted. Namely, the well known flows of the relativistic Toda
hierarchy (\ref{RTL+}), (\ref{RTL-}) have a much more rich dynamical content
as usually assumed. Even more is this true for the recently derived 
discretizations (\ref{dRTL+}), (\ref{dRTL-}) of these flows. Namely,
different parametrizations of the variables $(c,d)$ by canonically 
conjugated variables $(x,p)$ (corresponding to the bi--Hamiltonian structure
of the relativistic Toda hierarchy) allowed us to derive two new integrable
continuous time lattices and four new integrable discretizations, in addition
to the previously known ones. All the obtained systems are connected by means
of highly nontrivial B\"acklund transformations, which can be obtained by
identifying the variables $(c_k,d_k)$ for different models corresponding
to one and the same (continuous or discrete time) flow (\ref{RTL+}), 
(\ref{RTL-}), (\ref{dRTL+}), or (\ref{dRTL-}).

The methods of this paper can be used also in a more simple situation of
the usual Toda lattice, where they also lead to interesting findings. They
will be reported in a separate paper.

\end{document}